\documentclass[journal]{IEEEtran}
\usepackage{graphicx}
\usepackage{amsmath, amssymb}
\usepackage{algorithm}
\usepackage{algpseudocode}

\renewcommand{\baselinestretch}{0.96}    

\begin{document}
%
\title{A Learnt Half-Quadratic Splitting-Based Algorithm for Fast and High-Quality Industrial Cone-beam CT Reconstruction}
%
%
%

\author{Aniket Pramanik, Singanallur V. Venkatakrishnan, Obaidullah Rahman, Amirkoushyar Ziabari 
\thanks{This manuscript has been authored by UT-Battelle, LLC, under contract DE-AC05-00OR22725 with the US Department of Energy (DOE). Research sponsored by the US Department of Energy, Office of Energy Efficiency and Renewable Energy, Advanced Manufacturing Office and Technology Commercialization Fund (TCF-21-24881), under contract DE-AC05-00OR22725 with UT-Battelle, LLC.
The US government retains and the publisher, by accepting the article for publication, acknowledges that the US government retains a nonexclusive, paid-up, irrevocable, worldwide license to publish or reproduce the published form of this manuscript, or allow others to do so, for US government purposes. DOE will provide public access to these results of federally sponsored research in accordance with the DOE Public Access Plan (http://energy.gov/downloads/doe-public-access-plan).}}

\maketitle

\begin{abstract}
Industrial X-ray cone-beam CT (XCT) scanners are widely used for scientific imaging and non-destructive characterization. 
Industrial CBCT scanners use large detectors containing millions of pixels and the subsequent 3D reconstructions can be of the order of billions of voxels. 
In order to obtain high-quality reconstruction when using typical analytic algorithms, the scan involves collecting a large number of projections/views which results in large measurement times - limiting the utility of the technique. 
Model-based iterative reconstruction (MBIR) algorithms can produce high-quality reconstructions from fast sparse-view CT scans, but are computationally expensive and hence are avoided in practice. 
Single-step deep-learning (DL) based methods have demonstrated that it is possible to obtain fast and high-quality reconstructions from sparse-view data but they do not generalize well to out-of-distribution scenarios. 
In this work, we propose a half-quadratic splitting-based algorithm that uses convolutional neural networks (CNN) in order to obtain high-quality reconstructions from large sparse-view cone-beam CT (CBCT) measurements while overcoming the challenges with typical approaches.  
The algorithm alternates between the application of a CNN and a conjugate gradient (CG) step enforcing data-consistency (DC). 
The proposed method outperforms other methods 
on the publicly available Walnuts data-set.  
\end{abstract}

\begin{IEEEkeywords}
Convolutional neural network, Cone-beam CT, hybrid, regularization, deep-learning, iterative algorithm.
\end{IEEEkeywords}

%
\IEEEpeerreviewmaketitle

\section{Introduction}

\begin{figure}[t!]
\centering
\includegraphics[scale = 0.5,keepaspectratio=true,trim={8cm 5.5cm 2cm 5cm},clip]{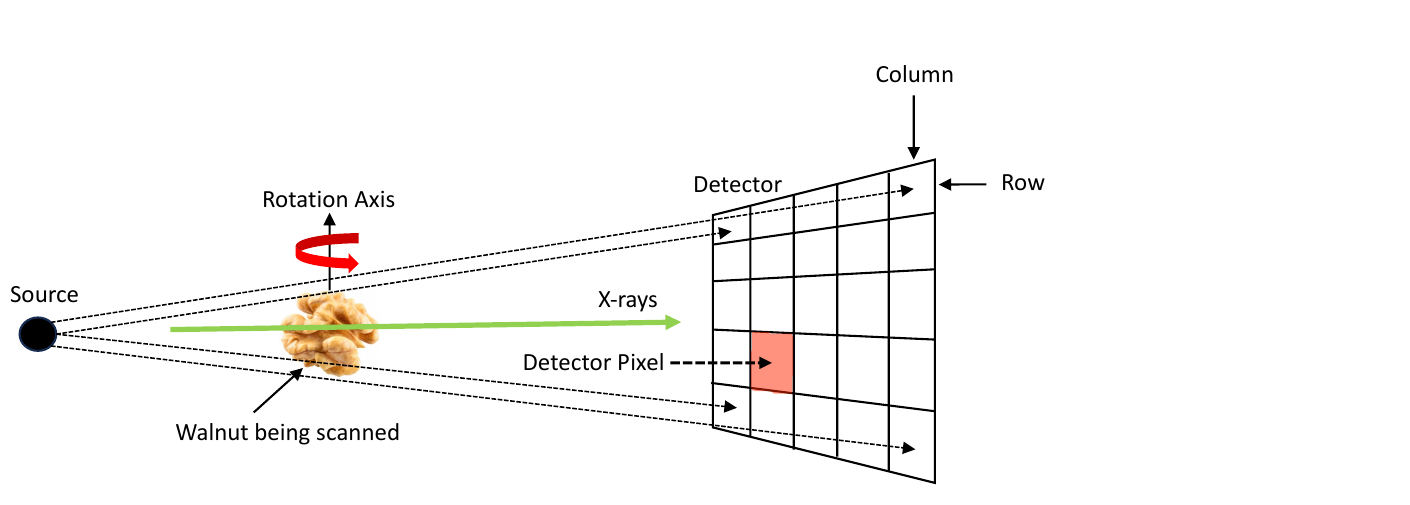}
\caption{The Cone-beam CT geometry. 
A cone-beam source of X-rays is incident on an object and the resulting transmitted signal is captured by a detector. 
The object is rotated and scanned at different angles (views).}
\label{fig:cbct}
\end{figure}

Industrial X-ray cone-beam CT (XCT) scanners are widely used for high-resolution scientific imaging and non-destructive characterization. 
In order to conduct a CT scan, an object is rotated about a single axis and multiple projections are acquired (see Fig.~\ref{fig:cbct}), that are typically processed by an analytic algorithm such as the Feldkamp-Davis-Kress (FDK) \cite{feldkamp1984practical} after appropriate pre-processing to obtain a 3D reconstruction. 
Unlike in medical CT, industrial XCT scanners use large detectors containing millions of pixels and the subsequent 3D reconstructions can be of the order of billions of voxels. 
A heuristic that is often used to obtain high-quality reconstruction 
is that the number of views needed is of the order of the number of columns of the detector, which can corresponds to total scan times of the order of several tens of minutes or even hours depending on the the material, desired reconstruction grid size and energy of the source. 

One way to accelerate the XCT scans is to acquire fewer projections, but naively processing them using the FDK algorithm results in reconstructions with significant artifacts. 
Model-based iterative reconstruction (MBIR) algorithms \cite{yu2010fast, thibault2007three}, that exploit the underlying imaging physics and edge-preserving models to regularize the image recovery,  have been developed and shown superior performance over analytical techniques for sparse-view cone-beam medical XCT. 
However, these techniques are computationally very expensive when dealing with large volumes as is typical for industrial cone-beam XCT which has made them impractical for routine use.

Single-stage deep learning (DL) approaches 
that utilize a convolutional neural network (CNN) to directly infer the final reconstruction from an initial analytic reconstruction \cite{jinUnet2017} have demonstrated that it is possible to obtain high-quality reconstructions while maintaining a lower computational complexity from sparse-view industrial XCT data \cite{ziabari2023enabling, rahman2023deep}.  
However, such methods require appropriate types of training data and do not generalize well to out of distribution scenarios such as varying measurement conditions. 
Furthermore, such methods lack any knobs to control image quality at test time in case there is any variation in imaging conditions.
Another class of algorithms developed recently are Deep Unrolled (DU) networks which combine the physics-based model with a data-driven DL model. 
These methods avoid overfitting and have shown reduced training data demands while improving performance \cite{xiang2021fista, su2022generalized}. 
However, these methods can require a large amount of memory that scales with the number of iterations - making them challenging for large scale industrial XCT.  
Anish et. al \cite{lahiri2023sparse} have proposed a memory-efficient DU approach consisting of a multi-iteration network alternating between a de-streaking CNN and the physics-based model and demonstrated its value for small data-sets. 

\begin{figure}[h!]%
\centering
\includegraphics[width=0.48\textwidth,keepaspectratio=true,trim={1.4cm 3.8cm 1.7cm 4.0cm},clip]{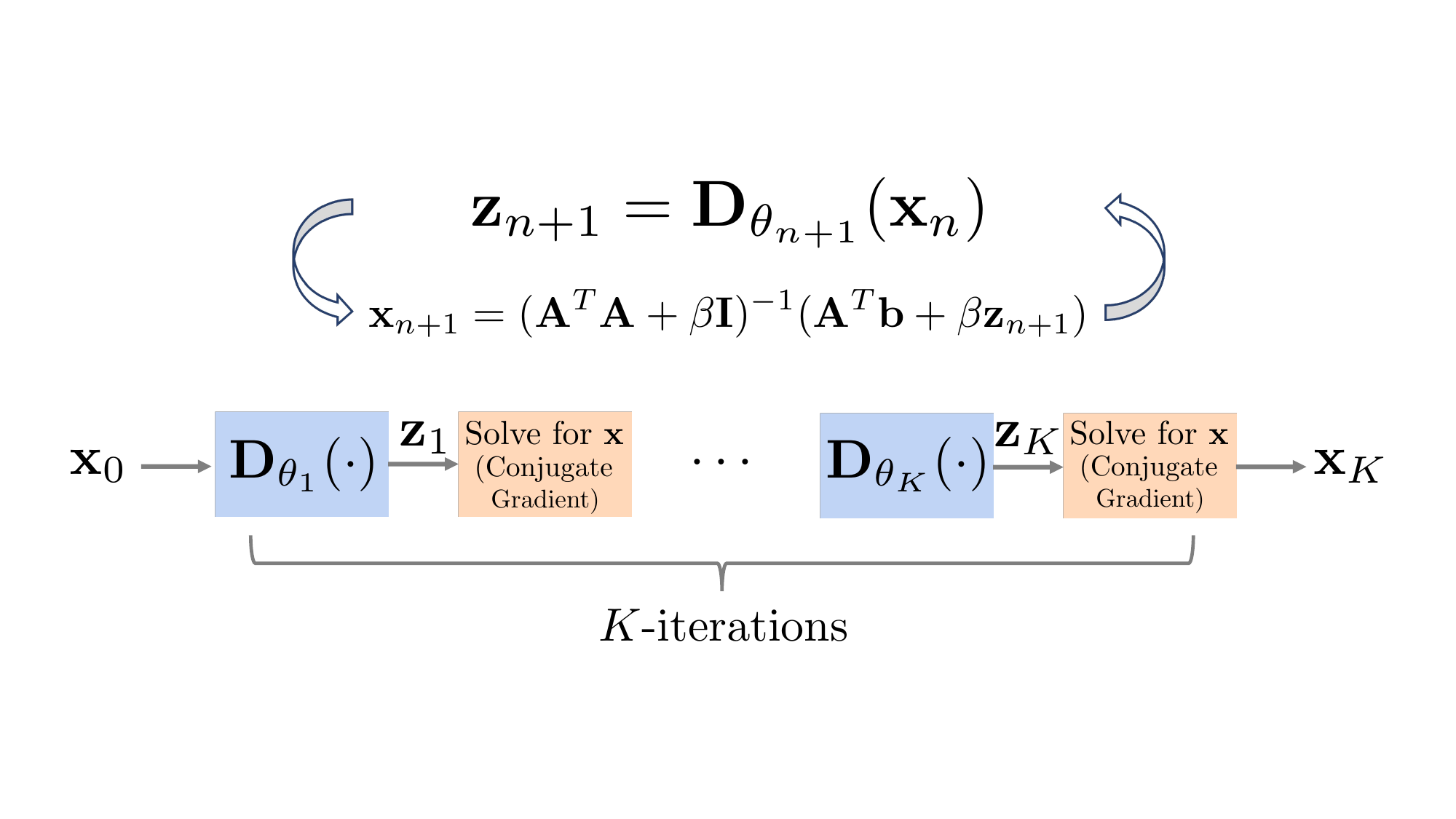}
\caption{Iterative rule of the proposed algorithm and its network architecture.
$\text{\bf{D}}$ represents a artifact suppression CNN and the block in orange represent a data-consistency block that ensures a good fit to the measurements.  
}%
\label{fig:proposed}%
\end{figure}

In this paper, we propose a learnt half-quadratic splitting algorithm that uses CNN to replace the proximal operation. 
The algorithm alternates for a fixed number of outer iterations between a CNN and the data-consistency (DC) block as shown in Fig. \ref{fig:proposed}.  
A regularization parameter controls the effect of both the blocks on reconstruction. 
The CNN essentially performs a denoising operation which is equivalent to solving for proximal operation for the regularizer. 
Unlike DU, it has a memory efficient implementation where CNN in each iteration is trained separately. 
We utilize a 2D CNN for training it with 2D patches obtained from the 3D volume images. 
We validate our method on the publicly available Walnuts\cite{der2019cone} cone-beam CT data-set - and demonstrate superior performance compared to the traditional MBIR and a single-stage U-Net approach. 
Overall, it reduces the time complexity significantly compared to MBIR. 
A key benefit of this memory efficient framework is its ability to be extended to large-scale image reconstruction problems without compromising on the performance.

\section{Theory}
MBIR methods have been widely developed for X-ray CT and obtain the reconstruction by minimizing a cost function
\begin{equation}
\label{cost}
\arg \min_{\mathbf x} \|\mathbf A \mathbf x - \mathbf b\|_2^2 + \lambda \mathbf Q(\mathbf x)
\end{equation}
where $\mathbf b \in \mathbb R^M$ is the log-normalized measurement vector, $\mathbf x \in \mathbb R^N$ is the 3D image to be recovered, $\mathbf A \in \mathbb R^{M \times N}$ is the cone-beam forward projector, $\mathbf Q(\cdot)$ is a regularization function that enforces certain desirable properties in the solution, and $\lambda$ balances the effect of regularizer  and the data-consistency (DC) term. 
Several algorithms such as optimized gradient methods \cite{fesslerogm} and iterative coordinate descent \cite{yu2010fast, thibault2007three} have been developed to find a solution to \eqref{cost} when using differentiable regularizers, but these are computationally very expensive when dealing with large data-sets in industrial XCT.  

Instead, we propose to use a different approach that uses a fixed number of outer iterations of the half-quadratic splitting (HQS) method and replace certain blocks with learnable functions. 
The HQS for solving \eqref{cost} results in a algorithm that alternates between the following two steps, 
\begin{eqnarray}
\mathbf z_{n+1} = \arg \min_{\mathbf z} \lambda \mathbf Q(\mathbf z) + \frac{\beta}{2} \|\mathbf x_n - \mathbf z\|_2^2 \label{pr} \\
\mathbf x_{n+1} = \arg \min_{\mathbf x} \frac{1}{2}\|\mathbf A \mathbf x - \mathbf b\|_2^2 + \frac{\beta}{2}\|\mathbf x - \mathbf z_{n+1}\|_2^2. \label{img}
\end{eqnarray}
where $\beta$ is a penalty parameter. 
The solution to the sub-problems in \eqref{pr}, \eqref{img} alternates between a DC step and a proximal operation,
\begin{eqnarray}
\mathbf x_{n+1} = (\mathbf A^T \mathbf A + \beta \mathbf I)^{-1}(\mathbf A^T \mathbf b + \beta \mathbf z_{n+1}) \\
\mathbf z_{n+1} = \rm prox_{\frac{\lambda}{\beta} \mathbf Q}(\mathbf x_n). \label{prox} 
\end{eqnarray}
The step in \eqref{prox} is a proximal operation for the regularizer $\mathbf Q(\cdot)$. 
It is equivalent to solving a Gaussian denoising problem whose maximum a posteriori (MAP) estimate is equivalent to solving for \eqref{pr}.

In the spirit of methods like plug-and-play priors (PnP) \cite{PnPVenkat13} we  replace the denoising step in \eqref{prox} with a CNN $\mathbf D_{\theta}(\cdot): \mathbb R^{\rm N} \to \mathbb R^{\rm N}$ with weights $\theta$. 
Therefore, the sub-problems \eqref{pr} and \eqref{img} are solved approximately using the following equations,

\begin{eqnarray}
\mathbf z_{n+1} = \mathbf D_{\theta_{n+1}}(\mathbf x_n) \label{cnn} \\
\mathbf x_{n+1} = (\mathbf A^H \mathbf A + \beta \mathbf I)^{-1}(\mathbf A^H \mathbf b + \beta \mathbf z_{n+1}). \label{dc} 
\end{eqnarray}
The network architecture is shown in Fig. \ref{fig:proposed}. 
The CNN $\mathbf D_{\theta}$ can have either shared or unshared weights across iterations i.e we can train a collection of CNNs  - one for each outer iteration. 
In practice, the DC step in \eqref{dc} is solved approximately using a fixed number of conjugate gradient (CG) algorithm iterations. 
The optimization steps in \eqref{cnn}, \eqref{dc} are essentially solving for,
\begin{equation}
\mathbf x_{n+1} = \arg \min_{\mathbf x} \frac{1}{2}\|\mathbf A \mathbf x - \mathbf b\|_2^2 + \frac{\beta}{2}\|\mathbf x - \mathbf D_{\theta_{n+1}}(\mathbf x_n)\|_2^2. \label{opt}   
\end{equation}
The regularization parameter $\beta$ in \eqref{opt} is chosen empirically and kept fixed across the iterations during training. $\beta$ balances between the denoising effect of the CNN and data-consistency. $\beta$ acts like a knob that can be varied during inference for unseen imaging conditions.

\begin{figure*}[t!]
\centering
\includegraphics[width =\textwidth,keepaspectratio=true,trim={1.5cm 10.5cm 1.5cm 11cm},clip]{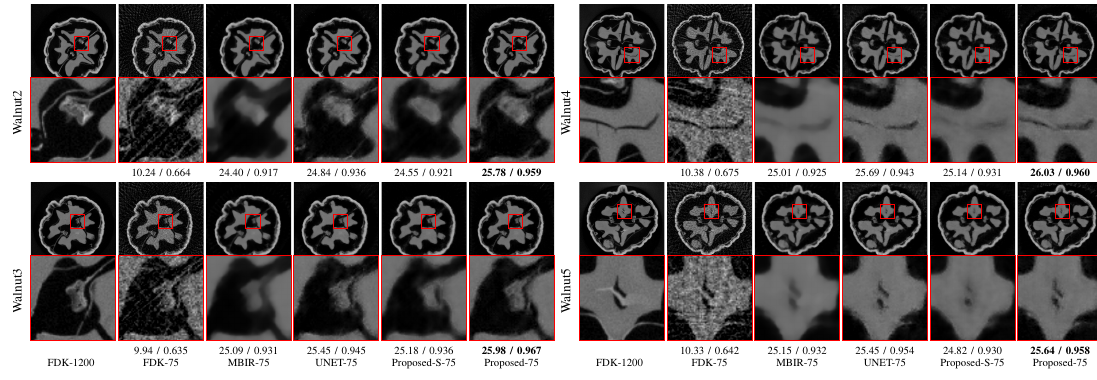}
\caption{Performance comparisons on Walnut Data. 
Walnut1 is the training data and the rest are testing data. 
Proposed-S-75 is the shared weights version of the algorithm applied on 75 views - a subsampling factor of 16 compared to the original measurements. 
PSNR in dB and SSIM values are reported for the zoomed patches. 
The proposed method with unshared weights preserves edges and details better compared to rest of the methods.}
\label{fig:perf_comp}
\end{figure*}

\begin{figure}[t!]
\centering
\includegraphics[scale = 0.7,keepaspectratio=true,trim={1.4cm 8.9cm 10.5cm 9.2cm},clip]{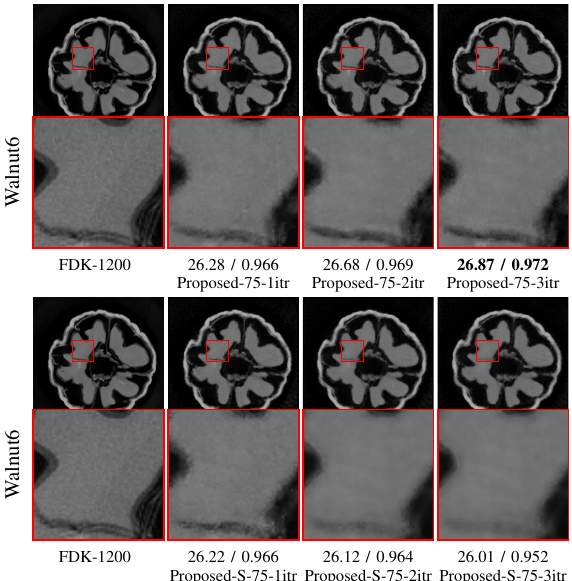}
\caption{Performance comparison of the two versions of the proposed method, shared vs unshared CNN weights. Each row displays results from three iterations. Proposed-S-75-$K$itr stands for the shared weights version on 75 views at $K^{\rm th}$ iteration while Proposed is the unshared one. Proposed-S is missing details due to excessive smoothing while the edges get sharper with more iterations for unshared weights. PSNR in dB and SSIM values are reported for the zoomed patches.}
\label{fig:perf_boi_sus}
\end{figure}

\begin{figure}[t!]
\centering
\includegraphics[scale = 0.7,keepaspectratio=true,trim={1.5cm 9cm 12.6cm 9.4cm},clip]{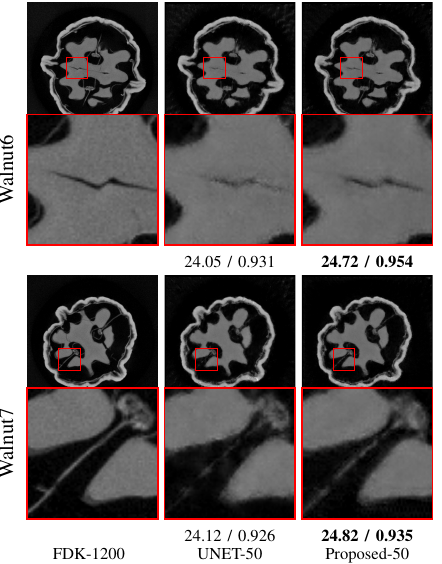}
\caption{Performance comparison of U-Net and the proposed method with unshared weights on out-of-distribution data as characterized by a different sparse-view measurement scenario. 
PSNR in dB and SSIM are reported for the zoomed patches. 
The results are shown on data from different sub-sampling scenario (50 views) which the networks have not been trained with. 
The proposed method shows significantly improved performance both visually and also in terms of performance metrics.}
\label{fig:perf_u_vs_p}
\end{figure}

\subsection{Memory Efficient Training and Implementation}
We propose to train the CNNs in each iteration separately. 
We consider CNNs with 2D convolution layers and therefore split the 3D CBCT data into several 2D patches for training. 
This approach reduces GPU memory consumption. 
During inference, the 3D volume image is obtained through the CNN in 2D slice-by-slice fashion. 
The CNNs from subsequent iterations are trained in a similar fashion with the data reconstructed from the previous iteration. 
The training is parallelized across multiple GPUs along the batch dimension of the training data. The DC step is implemented using the CBCT forward operator from the ASTRA Toolbox \cite{van2015astra} which can also be distributed over multiple GPUs. 
The training loss is the mean squared error (MSE) with expectation over the training samples,
\begin{equation}
\theta_{n+1} = \arg \min_{\theta} \mathbb E \bigg [\|\mathbf D_{\theta}(\mathcal P(\mathbf x_n)) - \mathcal P(\mathbf x_{\rm MBIR/FDK})\|_2^2 \bigg ] \label{loss}     
\end{equation}
where ground truth is either MBIR ($\mathbf x_{\rm MBIR}$) or FDK ($\mathbf x_{\rm FDK}$) reconstruction on the corresponding dense-view data, and $\mathcal P(\cdot)$ is a 2D patch extraction operator. 
The pseudo-code for the algorithm is shown in Algorithm \ref{Algo:training}.  
The variable $\rm ep$ denotes training epoch, $n$ is for iteration number and $i$ ranges over $N_t$ training samples.

\begin{algorithm}
	\caption{: Training: input=training data $\mathbf x_i; i=1,..,N_t$}
	 \label{Algo:training}
	\begin{algorithmic}
		\For {ep = $1,2,\ldots$ }
                \For {$n = 1,2,\ldots, K$}
                    \For {$i=1,2,\ldots, N_t$ }
				\State Extract 2D patches from $\mathbf x_{i,n}$ using $\mathcal P$
                    \State Train $\mathbf D_{\theta_{n+1}}(\cdot)$ using the loss function in \eqref{loss}
                    \State Compute $\mathbf z_{i,n+1} = \mathbf D_{\theta_{n+1}}(\mathbf x_{i,n})$.
                    \State Compute $\mathbf x_{i,n+1}$ from $\mathbf z_{i,n+1}$ using \eqref{dc}.
                    \EndFor
			\EndFor
		\EndFor
	\end{algorithmic} 
\end{algorithm}

Considering $K$ as number of iterations, the proposed method and MBIR have a time and space complexity of the order of $\mathcal O(K)$ and $\mathcal O(1)$ respectively. DU methods instead perform end-to-end training, and therefore, have a complexity of $\mathcal O(K)$ for both time and space. 
DL has the best possible complexity of $\mathcal O(1)$ for both. 
Therefore, our proposed approach reduces space complexity by a factor of $K$ over DU. 

\section{Experiments and Results}

\subsection{Data-set, CNN Architecture and Methods for Comparisons}
Experiments have been conducted on the publicly available CBCT Walnut data \cite{der2019cone}.
The detector size is 972 $\times$ 768 pixels and subsequently each scan consists of projection data from 1200 views (angles). 
In order to evaluate our algorithm on sparse-view data, we sub-sampled the views by factor of 16 to obtain sparse-view data (75 views) - simulating more than an order of magnitude faster scan time than the typical rate. 
The ground truth for training is an FDK reconstruction with 1200 views. 
The reconstructed volume has dimensions 450 $\times$  380 $\times$  380 voxels.  
The training data only includes a single scan (Walnut1) and all the methods are tested on six different scans (Walnut2-7).

We consider $K=3$ iterations for the proposed method. A full-size U-Net with four pooling and unpooling layers is chosen as the CNN for each iteration. It consists of 3 x 3 filters with the initial number of output channels being 64. The CNNs are trained with 2D patches of size 256 x 256 and the batch-size is set as 64. The training is parallelized across four 16GB Nvidia P100 GPUs. The CNN parameters are optimized using the Adam Optimizer for 500 epochs at a learning rate of $10^{-4}$. The inference was performed on each 2D slice sequentially to obtain the 3D volume image. The DC block consists of the CG algorithm operating on the 3D image generated by the CNN. The CG algorithm is run for ten iterations. The ASTRA forward operator is distributed over the four GPUs. We empirically choose the regularization parameter $\beta = 5 \times 10^{-2}$ and keep it fixed across all the iterations during training.

We compare the proposed method with unshared weights against FDK \cite{feldkamp1984practical}, MBIR \cite{thibault2007three,pyMBIR}, U-Net \cite{ronneberger2015u} and its shared weights version. 
Both U-Net and the proposed method are fed with FDK reconstruction on the sparse-view data. 
We also compare the iteration-by-iteration reconstruction quality of both the versions of the proposed method. 
MBIR has been implemented using pyMBIR \cite{pyMBIR} with ASTRA \cite{van2016fast}.
\subsection{Performance and Run-time Comparisons}
The reconstruction performances of various methods are shown in Fig. \ref{fig:perf_comp}. 
The PSNR and SSIM values are reported for the zoomed patches. The proposed method with unshared weights outperforms all the other methods. 
The one with shared weights (Proposed-S) loses key details due to excessive smoothing. 
The single-stage U-Net reconstructions are slightly inferior to the proposed due to overfitting errors and lack of measurement information through DC. 
MBIR gets rid of streak artifacts but loses some details due to blurring.

We compare the reconstruction quality at various iterations of the proposed method in Fig. \ref{fig:perf_boi_sus}. 
Proposed-S is the method with shared weights and its performance drops with iterations due to aggressive denoising (over-smoothing) at subsequent iterations. 
The one with unshared weights show sharper edges with iterations due to CNNs weights being optimal for other iterations. 
The PSNR and SSIM metrics drop for Proposed-S with iterations while they improve for the unshared weights. 

Fig. \ref{fig:perf_u_vs_p} displays the performance of U-Net and the proposed method with unshared weights on out-of-distribution data. We performed reconstructions on 50 views instead of 75 and found the proposed method to be outperforming U-Net visually as well as in terms of PSNR and SSIM metrics. The results validate the robustness of the proposed method on imaging conditions unseen by the network during training. 

We report the average run-time of each method in Table \ref{tab:runtime}. 
MBIR requires approximately two hundred iterations and therefore it is the slowest. 
The proposed method requires only three outer iterations and significantly reduces run-time. 
U-Net is much faster compared to the other two due to the absence of CG step.

\begin{table}[ht!]
\fontsize{7}{10}
\selectfont
\centering
\begin{tabular}{|cccc|}
\hline
MBIR & U-Net & Proposed & FDK\\  \hline 
1440 & 5 & 30 & 4 \\ \hline
\end{tabular}
\vspace{1em}
\caption{Average run-time reported in seconds for a volume of size $450 \times 380 \times 380$ voxels from measurements containing 75 views each of size $962 \times 768$ pixels.  
The proposed method is slower than the single-stage deep-learning method (U-Net) but is more than an order of magnitude faster than MBIR.}
\label{tab:runtime} 
\end{table}

\section{Conclusion}
We introduced a learnt half-quadratic splitting-based algorithm for CBCT reconstruction from sparse measurements. 
The proposed algorithm is a hybrid approach that uses CNN as a regularizer and a physics-based data consistency term; offering a memory-efficient implementation for large scale problems. 
It is significantly faster than the traditional MBIR approach, requiring very few iterations and also reduces over-fitting issues often dealt by DL methods such as U-Net. 
Experiments show that the proposed algorithm outperforms the state-of-the-art methods when validated on a publicly available CBCT data. 

\bibliographystyle{IEEEtran}
\bibliography{main}
\end{document}